\documentclass[11pt]{article}
\usepackage[russian]{babel}
\usepackage{amssymb,amsmath,amsthm}
\usepackage{graphicx}
\usepackage{epstopdf}
\newcommand{\bc}{\begin{center}}
\newcommand{\ec}{\end{center}}
\newcommand{\be}{\begin{equation}}
\newcommand{\ee}{\end{equation}}
\newcommand{\bea}{\begin{eqnarray}}
\newcommand{\eea}{\end{eqnarray}}
\newcommand{\ba}{\begin{array}}
\newcommand{\ea}{\end{array}}
\newcommand{\edc}{\end{document}}

\def\l{\lambda}

\begin{document}
\sloppy
УДК 517.98

\begin{center}
\textbf{\Large {Трансляционно-инвариантные и слабо периодические меры Гиббса для НС-моделей на дереве Кэли}}\
\end{center}

\begin{center}
Р.М.Хакимов\footnote{Институт математики, ул. Дурмон йули, 29, Ташкент, 100125, Узбекистан.\\
E-mail: rustam-7102@rambler.ru}
\end{center}

В данной статье изучаются Hard-Core (HC) модели на дереве Кэли. Для HC-модели с двумя состояниями улучшен один из результатов из работы \cite{KhR}: доказано, что при некоторых условиях на параметры существуют ровно две слабо периодические (не периодические) меры Гиббса. Кроме того, рассмотрены плодородные HC-модели с параметром активности $\lambda>0$ и четырьмя состояниями. Известно, что существуют три типа таких моделей. Обобщены некоторые результаты из работы \cite{XR2} и для одной из моделей доказана не единственность трансляционно-инвариантной меры Гиббса.\\

\textbf{Ключевые слова}: дерево Кэли, конфигурация, НС-модель, плодородный граф,
мера Гиббса, слабо периодические меры, трансляционно-инвариантные меры.\

\section{Введение.}\

Понятие меры Гиббса играет важную роль при изучении состояний физической системы, т.е. если мера Гиббса не единственна, то говорят, что существует фазовый переход, а это происходит, когда физическая система меняет свое состояние при изменении температуры (\cite {6}). Определение меры Гиббса и других понятий, связанных с теорией мер Гиббса, можно найти, например, в работах (\cite {6}-\cite {Rb}).

HC-модель на $d$-мерной решетке $ \mathbb Z ^ d$ была введена и изучена Мазелью и Суховым в работе \cite{Maz}. Работы (\cite {7}-\cite {KhR}) посвящены к изучению мер Гиббса для HC-модели с двумя состояниями на дереве Кэли. В работе (\cite {7}) была доказана единственность трансляционно-инвариантной меры. В работе \cite{RKh} изучены слабо периодические меры Гиббса для
HC-модели для нормального делителя индекса два и при некоторых
условиях на параметры показана единственность
(трансляционно-инвариантность) слабо периодической меры Гиббса, а
в работе \cite{XR} показана единственность
(трансляционно-инвариантность) слабо периодической меры Гиббса для
HC-модели при любых значениях параметров. В \cite{KhR} доказывается существование слабо периодических (не периодических)
мер Гиббса для HC-модели для нормального делителя индекса четыре на некоторых инвариантах при некоторых условиях
на параметры.

В работах (\cite {4}-\cite {RKh1}) изучены гиббсовские меры для HC-моделей с тремя состояниями на
дереве Кэли порядка $k\geq1$ и доказано, что
трансляционно-инвариантная мера Гиббса не единственна. В работе
\cite{bw} выделены плодородные HC-модели, соответствующие графам
"петля"\,, "свисток"\,, "жезл"\, и "ключ"\, в случае трех
состояний, и "палка"\,,\ "ключ"\ и "пистолет"\ в случае четырех
состояний. В работе \cite{KhR1} рассмотрены плодородные HC-модели с
четырьмя состояниями на дереве Кэли порядка два и в случаях
"палка"\,,\ "ключ"\ и "пистолет"\ доказана единственность
трансляционно-инвариантной меры Гиббса, а в \cite{XR2} эта задача изучена на дереве Кэли порядка два, три, четыре и пять. В случае "палка"\ найдены такие
критические значения параметра $\lambda$, что трансляционно-инвариантная мера не единственна.

Настоящая работа посвящена изучению слабо периодических мер Гиббса для HC-модели с двумя состояниями для нормального делителя индекса четыре и изучению  трансляционно-инвариантных мер Гиббса для плодородных HC-моделей с четырьмя состояниями,
соответствующих графам "палка"\, и "ключ"\ на дереве Кэли. В первом случае доказывается, что на некотором инвариантном множестве существуют \emph{ровно} две слабо периодические меры Гиббса, отличные от периодических. Во втором случае обобщается один из результатов работы \cite{XR2} в случае "палка". Кроме того, показана не единственность трансляционно-инвариантной меры Гиббса на дереве Кэли порядка $k\geq4$  в случае "ключ"\ и даны точные критические значения для параметра $\lambda$.

\section{Предварительные сведения.}

Дерево Кэли $\tau^k$ порядка $ k\geq 1 $ - бесконечное дерево, т.е. граф без циклов, из
каждой вершины которого выходит ровно $k+1$ ребер. Пусть
$\tau^k=(V,L,i)$, где $V$ --- есть множество вершин $\tau^k$, $L$
--- его множество ребер и $i$ --- функция инцидентности,
сопоставляющая каждому ребру $l\in L$ его концевые точки $x, y \in
V$. Если $i (l) = \{ x, y \} $, то $x$ и $y$ называются  {\it
ближайшими соседями вершины} и обозначается $l = \langle
x,y\rangle $. Расстояние $d(x,y), x, y \in V$ на дереве Кэли
определяется формулой
$$
d (x, y) = \min \{d | \exists x=x_0,x_1, \dots, x_{d-1},
x_d=y\in V \ \ \mbox {такие, что} \ \ \langle x_0,x_1\rangle,\dots, \langle x_
{d-1}, x_d\rangle\}.$$

Для фиксированного $x^0\in V$ обозначим $ W_n =\{x\in V | \ d (x, x^0) =n \}, \  V_n = \{x\in V | \ d (x, x^0) \leq n\}.$
Для $x\in W_{n}$ обозначим
$ S(x)=\{y\in{W_{n+1}}:d(x,y)=1\}.$

Известно, что существует взаимнооднозначное соответствие между
множеством $V$ вершин дерева Кэли порядка $k\geq 1$ и группой
$G_k$, являющейся свободным произведением $k+1$ циклических групп
второго порядка с образующими $a_1,...,a_{k+1}$, соответственно
(см. \cite{1}).

Рассмотрим HC-модель ближайших соседей с $m-$состояниями на
дереве Кэли. В этой модели каждой вершине $x$ ставится в
соответствие одно из значений $\sigma (x)\in \Phi=\{0,1,..,m\}$.
Значения $\sigma (x)=1,..,m$ означают, что вершина $x$ `занята', а
значение $\sigma (x)=0$ означает, что вершина $x$ `вакантна'.

Конфигурация $\sigma=\{\sigma(x),\ x\in V\}$ на дереве Кэли
задается как функция из $V$ в $\Phi$. Множество всех конфигураций
на $V$ обозначается через $\Omega$. Аналогичном образом можно
определить конфигурации в $V_n$ ($W_n$), и множество всех
конфигураций в $V_n$ ($W_n$) обозначается как $\Omega_{V_n}$
($\Omega_{W_n}$).

Рассмотрим множество $\Phi$ как множество вершин некоторого графа
$G$. С помощью графа $G$ определим $G-$допустимую конфигурацию
следующим образом. Конфигурация $\sigma$ называется
$G$-\textit{допустимой конфигурацией} на дереве Кэли (в $V_n$ или
$W_n$), если $\{\sigma (x),\sigma (y)\}$-ребро графа $G$ для любой
ближайшей пары соседей $x,y$ из $V$ (из $V_n$). Обозначим
множество $G$-допустимых конфигураций через $\Omega^G$
($\Omega_{V_n}^G$).

Множество активности \cite{bw} для графа $G$ есть функция $\l:G
\to R_+$ из множества $G$ во множество положительных
действительных чисел. Значение $\l_i$ функции $\l$ в вершине
$i\in\{0,1,..,m\}$ называется ее ``активностью''.

Для данных $G$ и $\l$ определим гамильтониан $G-$HC-модели как
 $$H^{\lambda}_{G}(\sigma)=\left\{%
\begin{array}{ll}
     \sum\limits_{x\in{V}}{\log \lambda_{\sigma(x)},} \ \ \ $ если $ \sigma \in\Omega^{G} $,$ \\
   +\infty ,\ \ \ \ \ \ \ \ \ \  \ \ \ $  \ если $ \sigma \ \notin \Omega^{G} $.$ \\
\end{array}%
\right. $$

\textbf{Определение 1.}(\cite{bw}) Граф называется плодородным,
если существует набор активности $\lambda$ такой, что
соответствующий гамильтониан имеет не менее двух
трансляционно-инвариантных мер Гиббса.

В этой работе рассматриваются случаи $m=1$ и $m=3$. При $m=3$ рассмотрим случай $\lambda_0=1, \
\lambda_1=\lambda_2=\lambda_3=\lambda \ $ и изучим соответствующие
этому случаю трансляционно-инвариантные меры Гиббса, а при $m=1$ изучим слабо периодические меры Гиббса.

Для HC-модели с четырьмя состояниями рассмотрим два типа плодородных (fertile) графов с четырьмя
вершинами $0,1,2,3$ (на множестве значений $\sigma(x)$), которые
имеют следующие виды (см.\cite{bw}-\cite{XR2}):

\[
\begin{array}{ll}
\mbox{\it палка}: &  \{0,1\}, \{1,2\}, \{2,3\};\\
\mbox{\it ключ}: &  \{0,1\}, \{0,2\}, \{1,2\}, \{2,3\}.
\end{array} \]
\begin{center}
\includegraphics[width=12cm]{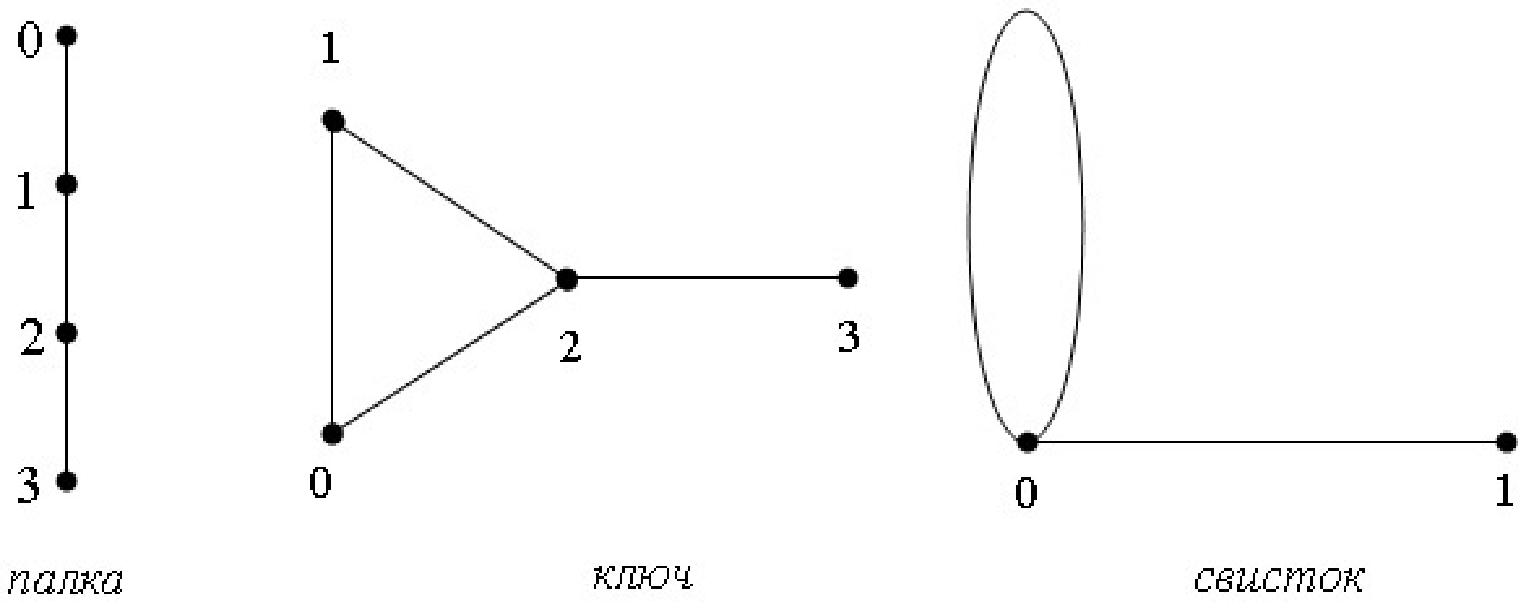}
\end{center}

В случае HC-модели с двумя состояниями в качестве $G$ имеем (см.\cite {7}-\cite {KhR}):
$$\textit{свисток}:   \{0,0\}, \{0,1\}.$$

Будем писать $x<y,$ если путь от $x^0$ до $y$ проходит через $x$.
Вершина $y$ называется прямым потомком $x$, если $y>x$ и $x, \ y$
являются соседями. Заметим, что в $\Im^k$ всякая вершина $x\neq x^0$
имеет $k$ прямых потомков, а вершина $x^0$ имеет $k+1$ потомков.

Для $\sigma_n\in\Omega_{V_n}^G$ положим
$$\#\sigma_n=\sum\limits_{x\in V_n}{\mathbf 1}(\sigma_n(x)\geq 1)$$
число занятых вершин в $\sigma_n$.

Пусть $z:x\mapsto z_x=(z_{0,x}, z_{1,x}, z_{2,x}, z_{3,x}) \in
R^4_+$ ($z:x\mapsto z_x=(z_{0,x}, z_{1,x}) \in R^2_+$ в случае $m=1$) векторнозначная функция на $V$. Для $n=1,2,\ldots$ и $\l>0$
рассмотрим вероятностную меру $\mu^{(n)}$ на $\Omega_{V_n}^G$,
определяемую как
\begin{equation}\label{rus2.1}
\mu^{(n)}(\sigma_n)=\frac{1}{Z_n}\lambda^{\#\sigma_n} \prod_{x\in
W_n}z_{\sigma(x),x}.
\end{equation}
Здесь $Z_n$-нормирующий делитель:
$$
Z_n=\sum_{{\widetilde\sigma}_n\in\Omega^G_{V_n}}
\lambda^{\#{\widetilde\sigma}_n}\prod_{x\in W_n}
z_{{\widetilde\sigma}(x),x}.
$$
Говорят, что вероятностная мера $\mu^{(n)}$ является
согласованной, если $\forall$ $n\geq 1$ и
$\sigma_{n-1}\in\Omega^G_{V_{n-1}}$:
\begin{equation}\label{rus2.2}
\sum_{\omega_n\in\Omega_{W_n}}
\mu^{(n)}(\sigma_{n-1}\vee\omega_n){\mathbf 1}(
\sigma_{n-1}\vee\omega_n\in\Omega^G_{V_n})=
\mu^{(n-1)}(\sigma_{n-1}).
\end{equation}
В этом случае существует единственная мера $\mu$ на $(\Omega^G,
\textbf{B})$ такая, что для всех $n$ и $\sigma_n\in
\Omega^G_{V_n}$
$$\mu(\{\sigma|_{V_n}=\sigma_n\})=\mu^{(n)}(\sigma_n),$$
где $\textbf{B}-$$\sigma$-алгебра, порожденная цилиндрическими
подмножествами $\Omega^G$.

\textbf{Определение 2.} Мера $\mu$, определенная формулой
(\ref{rus2.1}) с условием согласованности (\ref{rus2.2}),
называется ($G$-)HC-\textit{мерой Гиббса} с $\lambda>0$,
\textit{соответствующей функции} $z:\,x\in V
\setminus\{x^0\}\mapsto z_x$. Множество таких мер (для
всевозможных $z$) обозначается через ${\mathcal S}_G$. При этом
HC-мера Гиббса, соответствующая постоянной функции $z_x\equiv z$,
называется трансляционно-инвариантной.

\section{Слабо периодические меры Гиббса}

Гамильтониан HC-модели с двумя состояниями определяется по формуле
  $$H(\sigma)=\left\{%
\begin{array}{ll}
    J \sum\limits_{x\in{V}}{\sigma(x),} \ \ \ $ если $ \sigma \in\Omega $,$ \\
   +\infty ,\ \ \ \ \ \ \ \ \ \ $  \ если $ \sigma \ \notin \Omega $,$ \\
\end{array}%
\right. $$ где $J\in R$ и $\sigma (x)\in \Phi=\{0,1\}$.

Пусть $\widehat{G}_k-$ подгруппа группы $G_k$.

Если гиббсовская мера инвариантна относительно некоторой подгруппы
конечного индекса $ \widehat{G}_k\subset {G_k}$, то она называется
$\widehat{G}_k$ - периодической.

Известно \cite{7}, что каждой мере Гиббса для HC-модели на дереве
Кэли можно сопоставлять совокупность величин $z=\{z_x, x\in G_k
\},$ удовлетворяющих
\begin{equation}\label{rus2.3}
z_x=\prod_{y \in S(x)}(1+\lambda z_y)^{-1},
\end{equation} где $\lambda=e^{J_1}>0-$
параметр, $J_1=-J\beta$, $\beta={1\over T}$, $T>0-$температура.\

\textbf{Определение 3}. Совокупность величин $z=\{z_x,x\in G_k\}$
называется $ \widehat{G}_k$-периодической, если  $z_{yx}=z_x$ для
$\forall x\in G_k, y\in\widehat{G}_k.$\

$G_k-$ периодические совокупности называются трансляционно-инвариантными.

Для любого $x\in G_k $ множество $\{y\in G_k: \langle
x,y\rangle\}\setminus S(x)$ имеет единственный элемент, которого
обозначим через $x_{\downarrow}$ (см.\cite{5},\cite{8}).

Пусть $G_k/\widehat{G}_k=\{H_1,...,H_r\}$ фактор группа, где
$\widehat{G}_k$ нормальный делитель индекса $r\geq 1.$\

\textbf{Определение 4}. Совокупность величин $z=\{z_x,x\in G_k\}$
называется $\widehat{G}_k$ - слабо периодической, если
$z_x=z_{ij}$ при $x\in H_i, x_{\downarrow}\in H_j$ для $\forall
x\in G_k$.\

Заметим, что слабо периодическая совокупность $z$ совпадает с
обычной периодической (см. определение 3), если значение $z_x$ не
зависит от $x_{\downarrow}$.\

\textbf{Определение 5}. Мера $\mu$ называется
$\widehat{G}_k$-(слабо) периодической, если она соответствует
$\widehat{G}_k$-(слабо) периодической совокупности величин $z$.\

Пусть $A\subset\{1,2,...,k+1\}$ и $H_A=\{x\in G_k:\sum\limits_{i\in
A}w_x(a_i) -$четное число$ \}$, где $w_x(a_i)-$ число буквы $a_i$
в слове $x\in G_k$, ${G_k}^{(2)}=\{x\in G_k: \mid
x\mid-\mbox{четное число}\},$ где $\mid x\mid-$ длина слова $x\in
G_k$ и ${G_k}^{(4)}=H_A\cap{G_k}^{(2)}-$ нормальный делитель
индекса 4.\

Рассмотрим фактор-группу $G_k/{G_k}^{(4)}=\{H_0, H_1, H_2, H_3\},$ где
$$H_0=\{x\in G_k: \sum\limits_{i\in
A}w_x(a_i)-\mbox{четно}, |x|-\mbox{четно}\}$$
$$H_1=\{x\in G_k: \sum\limits_{i\in
A}w_x(a_i)-\mbox{нечетно}, |x|-\mbox{четно}\}$$
$$H_2=\{x\in G_k: \sum\limits_{i\in
A}w_x(a_i)-\mbox{четно}, |x|-\mbox{нечетно}\}$$
$$H_3=\{x\in G_k: \sum\limits_{i\in
A}w_x(a_i)-\mbox{нечетно}, |x|-\mbox{нечетно}\}$$ Тогда в силу (\ref{rus2.3})
$G_k^{(4)}-$ слабо периодическая совокупность величин $z_x$ имеет вид
$$
z_x=\left\{%
\begin{array}{ll}
    z_1, & {x \in H_3, \ x_{\downarrow} \in H_1} \\
    z_2, & {x \in H_1, \ x_{\downarrow} \in H_3} \\
    z_3, & {x \in H_3, \ x_{\downarrow} \in H_0} \\
    z_4, & {x \in H_0, \ x_{\downarrow} \in H_3} \\
    z_5, & {x \in H_1, \ x_{\downarrow} \in H_2} \\
    z_6, & {x \in H_2, \ x_{\downarrow} \in H_1} \\
    z_7, & {x \in H_2, \ x_{\downarrow} \in H_0} \\
    z_8, & {x \in H_0, \ x_{\downarrow} \in H_2}, \\
    \end{array}%
\right.$$ где $z_x$ удовлетворяет системе уравнений

\begin{equation}\label{rus2.4}\left\{\begin{array}{ll}
    z_{1}=\frac{1}{(1+\lambda z_4)^i}\cdot\frac{1}{(1+\lambda z_2)^{k-i}} \\
    z_{2}=\frac{1}{(1+\lambda z_6)^i}\cdot\frac{1}{(1+\lambda z_1)^{k-i}} \\
    z_{3}=\frac{1}{(1+\lambda z_4)^{i-1}}\cdot\frac{1}{(1+\lambda z_2)^{k-i+1}}\\
    z_{4}=\frac{1}{(1+\lambda z_3)^{i-1}}\cdot\frac{1}{(1+\lambda z_7)^{k-i+1}}\\
    z_{5}=\frac{1}{(1+\lambda z_6)^{i-1}}\cdot\frac{1}{(1+\lambda z_1)^{k-i+1}}\\
    z_{6}=\frac{1}{(1+\lambda z_5)^{i-1}}\cdot\frac{1}{(1+\lambda z_8)^{k-i+1}}\\
    z_{7}=\frac{1}{(1+\lambda z_5)^i}\cdot\frac{1}{(1+\lambda z_8)^{k-i}} \\
    z_{8}=\frac{1}{(1+\lambda z_3)^i}\cdot\frac{1}{(1+\lambda z_7)^{k-i}}. \\
\end{array}\right.
\end{equation}
Здесь $i=|A|-$мощность множества $A$. Из системы уравнений (\ref{rus2.4}) можно получить следующую систему уравнений (см. \cite{KhR})
\begin{equation}\label{rus2.5}\left\{\begin{array}{ll}
    z_{1}=\frac{(1+\lambda z_{7})^k}{((1+\lambda z_7)^{k/i}+\lambda z_8^{1-1/i})^i}\cdot\frac{1}{(1+\lambda z_2)^{k-i}} \\
    \\
    z_{2}=\frac{(1+\lambda z_8)^k}{((1+\lambda z_8)^{k/i}+\lambda z_7^{1-1/i})^i}\cdot\frac{1}{(1+\lambda z_1)^{k-i}} \\
    \\
    z_{7}=\frac{(1+\lambda z_1)^k}{((1+\lambda z_1)^{k/i}+\lambda z_2^{1-1/i})^i}\cdot\frac{1}{(1+\lambda z_8)^{k-i}} \\
    \\
    z_{8}=\frac{(1+\lambda z_2)^k}{((1+\lambda z_2)^{k/i}+\lambda z_1^{1-1/i})^i}\cdot\frac{1}{(1+\lambda z_7)^{k-i}} \\
\end{array}\right.
\end{equation}
Из работы \cite{KhR} известно, что множество $I_2=\{(z_1, z_2, z_7, z_8)\in R^4: z_1=z_7, \ z_2=z_8\},$
является инвариантным относительно отображения $W:R^4 \rightarrow R^4,$ определенным следующим образом:

$$
\left\{%
\begin{array}{ll}
    z_{1}^{'}=\frac{(1+\lambda z_{7})^k}{((1+\lambda z_7)^{k/i}+\lambda z_8^{1-1/i})^i}\cdot\frac{1}{(1+\lambda z_2)^{k-i}}
    \\[3mm]
    z_{2}^{'}=\frac{(1+\lambda z_8)^k}{((1+\lambda z_8)^{k/i}+\lambda z_7^{1-1/i})^i}\cdot\frac{1}{(1+\lambda z_1)^{k-i}}
    \\[3mm]
    z_{7}^{'}=\frac{(1+\lambda z_1)^k}{((1+\lambda z_1)^{k/i}+\lambda z_2^{1-1/i})^i}\cdot\frac{1}{(1+\lambda z_8)^{k-i}}
    \\[3mm]
    z_{8}^{'}=\frac{(1+\lambda z_2)^k}{((1+\lambda z_2)^{k/i}+\lambda z_1^{1-1/i})^i}\cdot\frac{1}{(1+\lambda z_7)^{k-i}} \\
\end{array}%
\right.
$$

Кроме того, в \cite{KhR} (см.Лемма 2) было доказано, что если на $I_2$ существуют слабо периодические меры Гиббса, то они являются либо трансляционно-инвариантными, либо слабо периодическими(не периодическими).

Перепишем систему (\ref{rus2.5}) на $I_2$ при $k=2,
i=1$
$$
\left\{%
\begin{array}{ll}
    z_{1}=\frac{(1+\lambda z_{1})^2}{(1+\lambda z_1)^{2}+\lambda} \cdot\frac{1}{1+\lambda z_2}
    \\[3mm]
    z_{2}=\frac{(1+\lambda z_2)^2}{(1+\lambda z_2)^{2}+\lambda }\cdot\frac{1}{1+\lambda z_1} \\
    \end{array}%
\right.
$$
После обозначений $x=1+\lambda z_1>1$ и $y=1+\lambda z_2>1$ из последней системы уравнений имеем

$$
\left\{%
\begin{array}{ll}
    x=f(y)    \\
    y=f(x),
    \end{array}%
\right.
$$ где $f(x)={\lambda x^2\over(x^2+\lambda)(x-1)}$. Уравнение
$f(x)=x$ имеет единственное положительное решение при любых
$\lambda>0$, т.к. уравнение
$$x={\lambda x^2\over(x^2+\lambda)(x-1)}=f(x)$$
эквивалентно уравнению $x^3-x^2-\lambda=0$, которое по теореме Декарта
о количестве положительных корней многочлена имеет не
более одного положительного решения и
$$f(1)=-\lambda, \ \lim_{x\rightarrow+\infty}f(x)=+\infty.$$

Очевидно, что это решение больше единицы. Кроме того, оно
находится среди решений уравнения $f(f(x))=x$. Поэтому рассмотрим
уравнение
$${x-f(f(x))\over x-f(x)}=0,$$
которое эквивалентно уравнению
$$h(x)=x^6-(\lambda+2)x^5+(5\lambda+1)x^4-\lambda(2\lambda+5)x^3+2\lambda(2\lambda+1)x^2-3\lambda^2x+\lambda^2=0.$$
Из этого уравнения имеем $h(1)=\lambda>0$ и $h(x)\rightarrow
+\infty$ при $x\rightarrow+\infty$. Кроме того, график функции
$h(x)$ касается оси Ox при $x=2, \lambda=4$, т.к.
$h(2)=-(\lambda-4)(5\lambda+4).$ Отсюда уравнение $h(x)=0$ не
имеет решений при $\lambda<4$, имеет одно решение при $\lambda=4$
и имеет по крайней мере два решения при $\lambda>4$.

\begin{center}
\includegraphics[width=6cm]{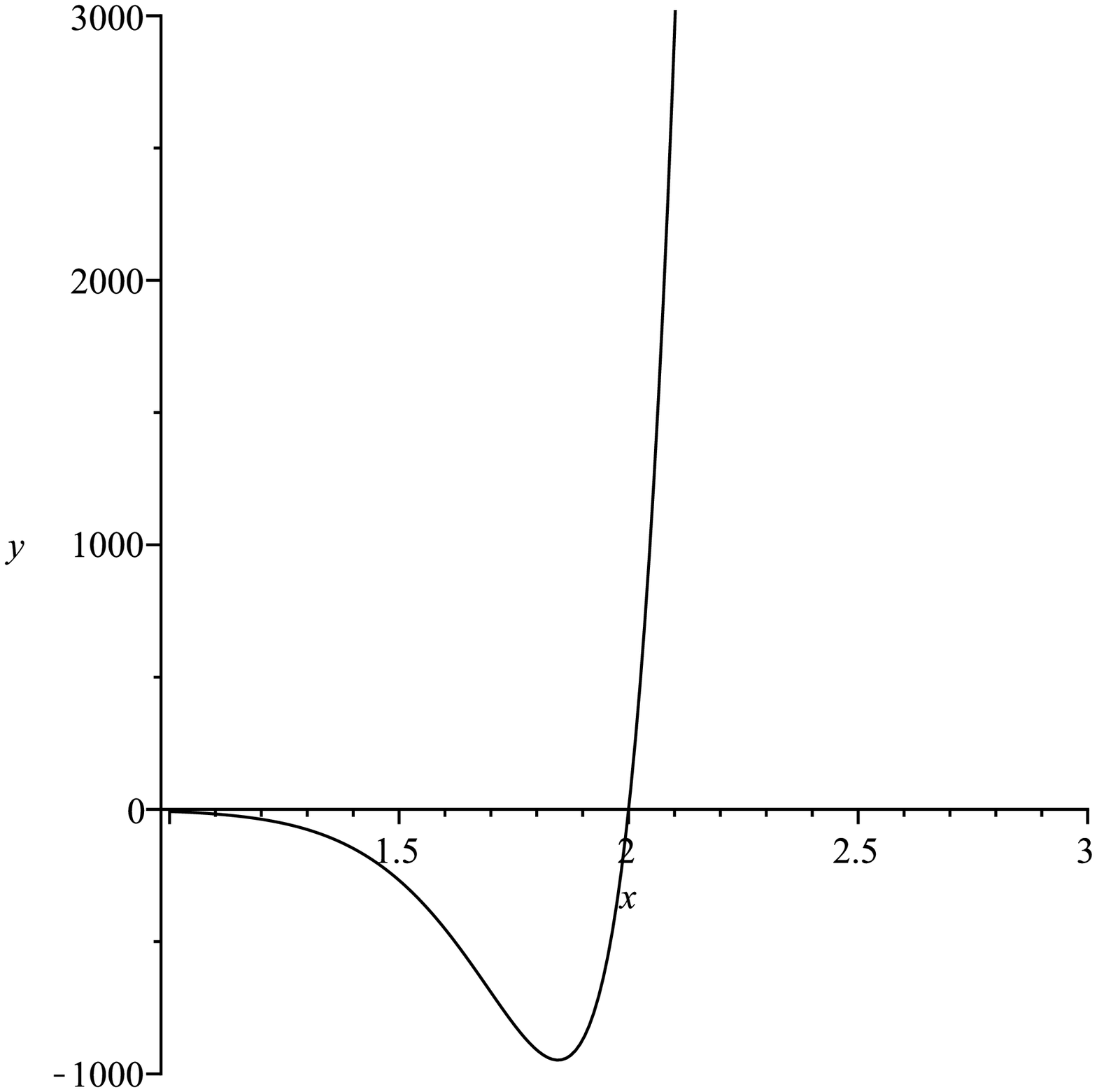} \includegraphics[width=6cm]{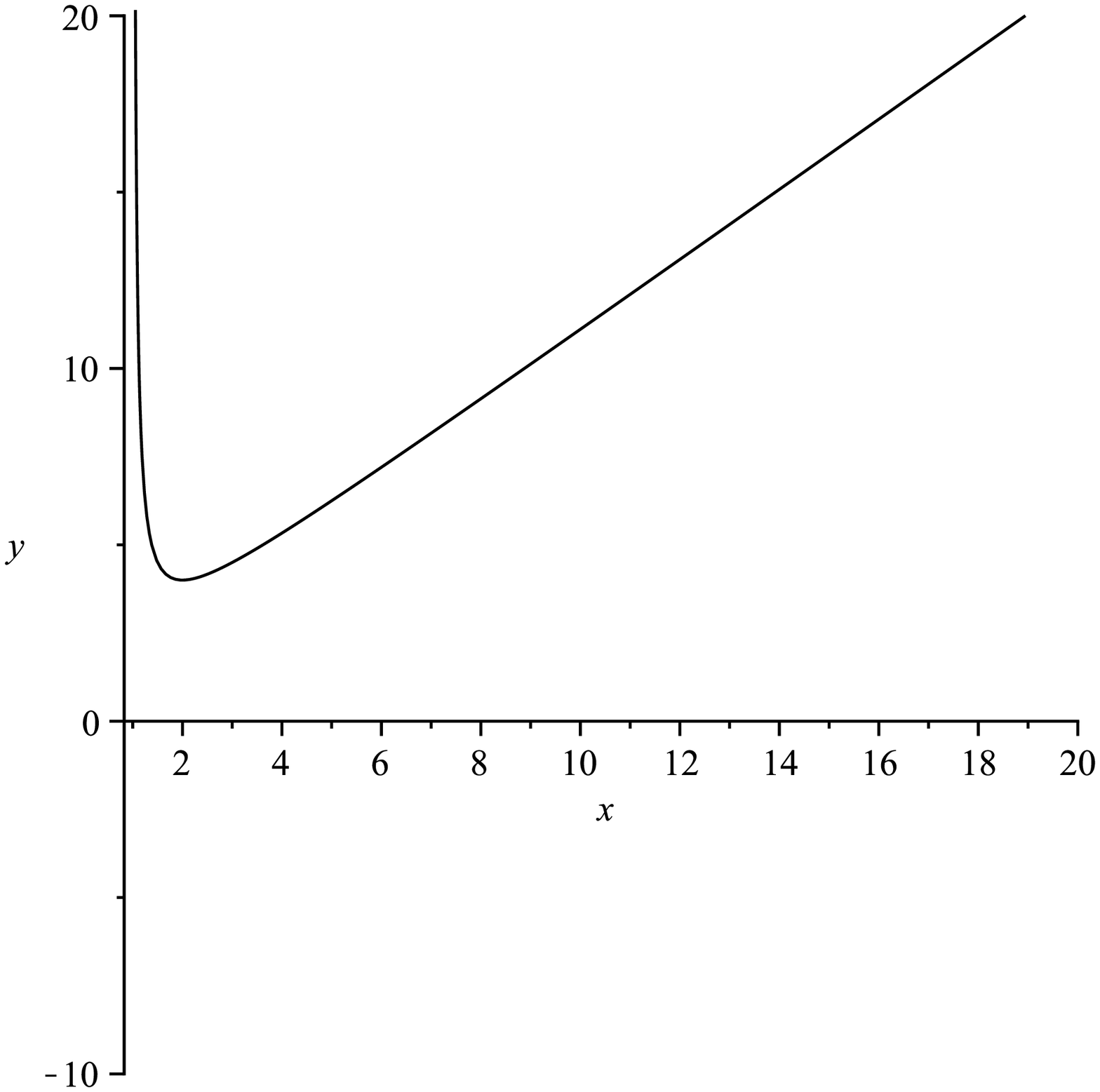}
\end{center}
\begin{center}{\footnotesize \noindent
Рис.~1.
График функции $\psi(x)$ при $x>1$ (слева). График функции $\varphi(x)$ при $x>1$ (справа).}
\end{center}\

Покажем, что уравнение $h(x)=0$ при $\lambda>4$ имеет ровно два решения. Оно имеет решение $x=x(\lambda)$. Но мы рассмотрим это уравнение относительно переменнной $\lambda$, т.е.
$$-(x-1)(2x^2-2x+1)\lambda^2-x^2(x^3-5x^2+5x-2)\lambda+x^4(x-1)^2=0$$
и получим решения вида
$$\lambda_{1_2}={x^2(x^3-5x^2+5x-2)\pm x^3\sqrt{x^4-2x^3+3x^2-2x+1}\over -2(x-1)(2x^2-2x+1)}.$$
Очевидно, что $\lambda_1<0$ и $\lambda_2>0$. Отсюда рассмотрим
$$\lambda=\lambda_2={x^2(x^3-5x^2+5x-2)- x^3\sqrt{x^4-2x^3+3x^2-2x+1}\over -2(x-1)(2x^2-2x+1)}=\varphi(x).$$
Рассмотрим уравнение
$$\varphi^{'}(x)={g_1(x)+g_2(x)\sqrt{D}\over g_3(x)\sqrt{D}}=0,$$
где
$$g_1(x)=2x^5(4x^7-18x^6+38x^5-52x^4+48x^3-31x^2+13x-3),$$
$$g_2(x)=-2x(4x^6-22x^5+52x^4-66x^3+50x^2-21x+4),$$
$$g_3(x)=4(2x^3-4x^2+3x-1)^2, \ D=x^6(x^4-2x^3+3x^2-2x+1).$$
Оно эквивалентно уравнению $(x-1)^3\psi(x)=0,$ где
$\psi(x)=64x^{12}-480x^{11}+1760x^{10}-4224x^9+7312x^8-9592x^7+9736x^6-7696x^5+4696x^4-2160x^3+712x^2-152x+16$. С помощью программы MapLE можно увидеть, что уравнение $\psi(x)=0$ имеет один вещественный корень $x_0=2$ при $x>1$ (Рис.1, слева), т.е. функция $\varphi(x)$ при $x>1$ имеет одну критическую точку $x_0=2$. Кроме того, $\varphi(x)\rightarrow+\infty$ при $x\rightarrow1$ и $x\rightarrow+\infty$. Следовательно, функция $\varphi(x)$ убывает при $1<x<2=x_{min}$ и возрастает при $x>2$ (Рис.1, справа). Из всего сказанного следует, что каждому значению $\lambda$ соответствуют только два значения $x$ при $\lambda>\lambda_{cr}$, где $\lambda_{cr}=\varphi(2)=4$.

Итак, справедлива следующая

\textbf{Теорема 1.} \textit{Пусть $k=2, i=1, \lambda_{cr}=4$. Тогда для HC-модели в случае нормального
делителя индекса четыре на $I_2$ при $\lambda<\lambda_{cr}$ существует одна слабо периодическая мера
Гиббса, которая является трансляционно-инвариантной, при
$\lambda=\lambda_{cr}$ существуют две слабо периодические меры
Гибсса, одна из которых является трансляционно-инвариантной,
другая слабо периодической (не периодической) и при
$\lambda>\lambda_{cr}$ существуют ровно три слабо периодические меры
Гибсса, одна из которых является трансляционно-инвариантной, остальные две слабо
периодическими (не периодическими).}\

\textbf{Замечание 1.} В работе \cite{KhR} было доказано, что при
$\lambda>\lambda_{cr}$ существуют не менее двух слабо
периодических (не периодических) мер Гибсса.\

\section{Трансляционно-инвариантные меры Гиббса}

Пусть $L(G)$-множество ребер графа $G$, обозначим через $A\equiv
A^G=\big(a_{ij}\big)_{i,j=0,1,2,3}$ матрицу смежности $G$, т.е.
\begin{equation}\label{rus2.3} a_{ij}\equiv a^G_{ij}=\left\{\begin{array}{ll}
1,\ \ \mbox{если}\ \ \{i,j\}\in L(G),\\
0, \ \ \mbox{если} \ \  \{i,j\}\notin L(G).
\end{array}\right.
\end{equation}
В следующей теореме сформулировано условие на $z_x$, гарантирующее
согласованность меры $\mu^{(n)}$.

\textbf{Теорема 2.}\cite{Rb} \label{rust1} Вероятностные меры
$\mu^{(n)}$, $n=1,2,\ldots$, заданные формулой (\ref{rus2.1}),
согласованны тогда и только тогда, когда для любого $x\in V$ имеют
место следующие равенства:
$$
z'_{0,x}=\lambda \prod_{y\in S(x)}{a_{00}z'_{0,y}+
a_{01}z'_{1,y}+a_{02}z'_{2,y}+a_{03}\over
a_{30}z'_{0,y}+a_{31}z'_{1,y}+a_{32}z'_{2,y}+a_{33}},$$
$$z'_{1,x}=\lambda \prod_{y\in S(x)}{a_{10}z'_{0,y}+
a_{11}z'_{1,y}+a_{12}z'_{2,y}+a_{13}\over
a_{30}z'_{0,y}+a_{31}z'_{1,y}+a_{32}z'_{2,y}+a_{33}},$$
$$z'_{2,x}=\lambda \prod_{y\in S(x)}{a_{20}z'_{0,y}+
a_{21}z'_{1,y}+a_{22}z'_{2,y}+a_{23}\over
a_{30}z'_{0,y}+a_{31}z'_{1,y}+a_{32}z'_{2,y}+a_{33}},$$ где
$z'_{i,x}=\lambda z_{i,x}/z_{3,x}, \ \ i=0,1,2$.\

Мы полагаем, что $z_{3,x}\equiv 1$ и $z_{i,x}=z'_{i,x}>0,\ \
i=0,1,2$. Тогда для любых функций $x\in V\mapsto
z_x=(z_{0,x},z_{1,x},z_{2,x})$, удовлетворяющих равенству

\begin{equation}\label{rus3.1}
z_{i,x}=\lambda \prod_{y\in S(x)}{a_{i0}z_{0,y}+
a_{i1}z_{1,y}+a_{i2}z_{2,y}+a_{i3}\over
a_{30}z_{0,y}+a_{31}z_{1,y}+a_{32}z_{2,y}+a_{33}}, \ \ i=0,1,2,
\end{equation}
существует единственная $G$-HC-мера Гиббса $\mu$ и наоборот.
Рассмотрим трансляционно-инвариантные решения, в которых $z_x=z\in
R^3_+$, $x\neq x_0$. В случаях $G=\textit{палка}$ и
$G=\textit{ключ}$ из (\ref{rus3.1})
получим следующие системы уравнений:
\begin{equation}\label{rus3.01} \left\{\begin{array}{ll}
z_0=\lambda\left({ z_1\over z_2}\right)^k,\\[2mm]
z_1=\lambda\left({z_0+z_2\over z_2}\right)^k,\\[2mm]
z_2=\lambda\left({z_1+1\over z_2}\right)^k,
\end{array}\right.
\end{equation}

\begin{equation}\label{rus3.02} \left\{\begin{array}{ll}
z_0=\lambda\left({ z_1+z_2\over z_2}\right)^k,\\[2mm]
z_1=\lambda\left({z_0+z_2\over z_2}\right)^k,\\[2mm]
z_2=\lambda\left({z_0+z_1+1\over z_2}\right)^k,
\end{array}\right.
\end{equation}
соответственно.

\textbf{Лемма.} (\cite{KhR1}) \textit{В системе уравнений (\ref{rus3.02}) $z_0=z_1.$}

\subsection{Случай $G=\textit{палка}$}\

В случае $G=\textit{палка}$ из третьего уравнения системы
(\ref{rus3.01}) найдем $z_2$ и подставим в первое уравнение.
Полученные выражения для $z_2$ и $z_0$ подставим во второе
уравнение. В результате получим уравнение ($z=z_1$)
\begin{equation}\label{rus3.5}
z=\lambda\cdot \left({z^k\over (z+1)^k}+1\right)^k=f(z).
\end{equation}
Из уравнения (\ref{rus3.5}) найдем $\lambda$
\begin{equation}\label{rus3.6}
\lambda=\varphi(z)={z\over \left({z^k\over
(z+1)^k}+1\right)^k}.
\end{equation}

\textbf{Замечание 2.} 1. В работе \cite{KhR1} при $k=2$ и $\lambda
>0$ было доказано, что существует только одна трансляционно-инвариантная мера Гиббса.

2. В работе \cite{XR2} при $k=3, \ k=4$ и $\lambda
>0$ было доказано, что существует только одна трансляционно-инвариантная мера Гиббса, а при $k=5$ были найдены приближенные значения для
$\lambda_{cr}^{(1)}\approx 0.8800478543, \ \lambda_{cr}^{(2)}\approx 1.078094055$ такие, что при $\lambda<\lambda_{cr}^{(1)}$ и $\lambda>\lambda_{cr}^{(2)}$ существует только одна трансляционно-инвариантная мера Гиббса, при $\lambda=\lambda_{cr}^{(1)}$ или $\lambda=\lambda_{cr}^{(2)}$ существуют ровно две трансляционно-инвариантные меры Гиббса и при $\lambda_{cr}^{(1)}<\lambda<\lambda_{cr}^{(2)}$ существуют ровно три трансляционно-инвариантные меры Гиббса.

Следующая теорема является обобщением Теоремы 2 из \cite{XR2}:

\textbf{Теорема 3.} \textit{Пусть $k\geq5$. Тогда в случае $G=\textit{палка}$ для Hard-Core модели существуют $\lambda_{cr}^{(1)}=\varphi(z_2), \ \lambda_{cr}^{(2)}=\varphi(z_1)$ такие, что}
\textit{при $\lambda<\lambda_{cr}^{(1)}$ и $\lambda>\lambda_{cr}^{(2)}$ существует только одна трансляционно-инвариантная мера Гиббса,}
\textit{при $\lambda=\lambda_{cr}^{(1)}$ или $\lambda=\lambda_{cr}^{(2)}$ существуют две трансляционно-инвариантные меры Гиббса и }
\textit{при $\lambda_{cr}^{(1)}<\lambda<\lambda_{cr}^{(2)}$ существуют только три трансляционно-инвариантные меры Гиббса, где $z_1, \ z_2-$точки максимума и минимума функции $\varphi(z)$, соответственно.}

\textbf{Доказательство.} Рассмотрим (\ref{rus3.6}) при $k\geq5$. Проанализируя функцию $\varphi(z),$ заметим, что
$\varphi(z)>0, \ \varphi(0)=0$ и $\varphi(z)\rightarrow +\infty$
при $z\rightarrow+\infty$. Ее производная имеет вид
$$\varphi'(z)={(z+1)^{k^2-1}[z^{k+1}+(1+z)^{k+1}+z^k-k^2z^k]\over [z^k+(z+1)^k]^{k+1}}.$$

По теореме Декарта многочлен $h(z)=z^{k+1}+(1+z)^{k+1}+z^k-k^2z^k$ имеет не более двух положительных корня. Очевидно, что $h(0)=1$ и $h(z)\rightarrow+\infty$ при $z\rightarrow+\infty$. Кроме того, $h(k+1)<0$ (Рис.2, слева). Следовательно, многочлен $h(z)$ имеет ровно два корня $z_1, \ z_2$, т.е. для функции $\varphi(z)$ существуют две критические точки. Тогда функция
$\varphi(z)$ возрастает при $0<z<z_1$ и $z>z_2$, убывает
при $z_1<z<z_2$ (Рис.2, справа). Отсюда $z_1=z_{max}, \
z_2=z_{min}$. А это значит, что каждому значению $\lambda>0$
соответствует только одно значение $z$ при
$\lambda<\varphi(z_2)=\lambda_{cr}^{(1)}$ и
$\lambda>\varphi(z_1)=\lambda_{cr}^{(2)}$, два значения при
$\lambda=\lambda_{cr}^{(1)}$ или $\lambda=\lambda_{cr}^{(2)}$ и
ровно три значения при $\lambda_{cr}^{(1)}<\lambda<\lambda_{cr}^{(2)}$.
Теорема доказана.\

\begin{center}
\includegraphics[width=6cm]{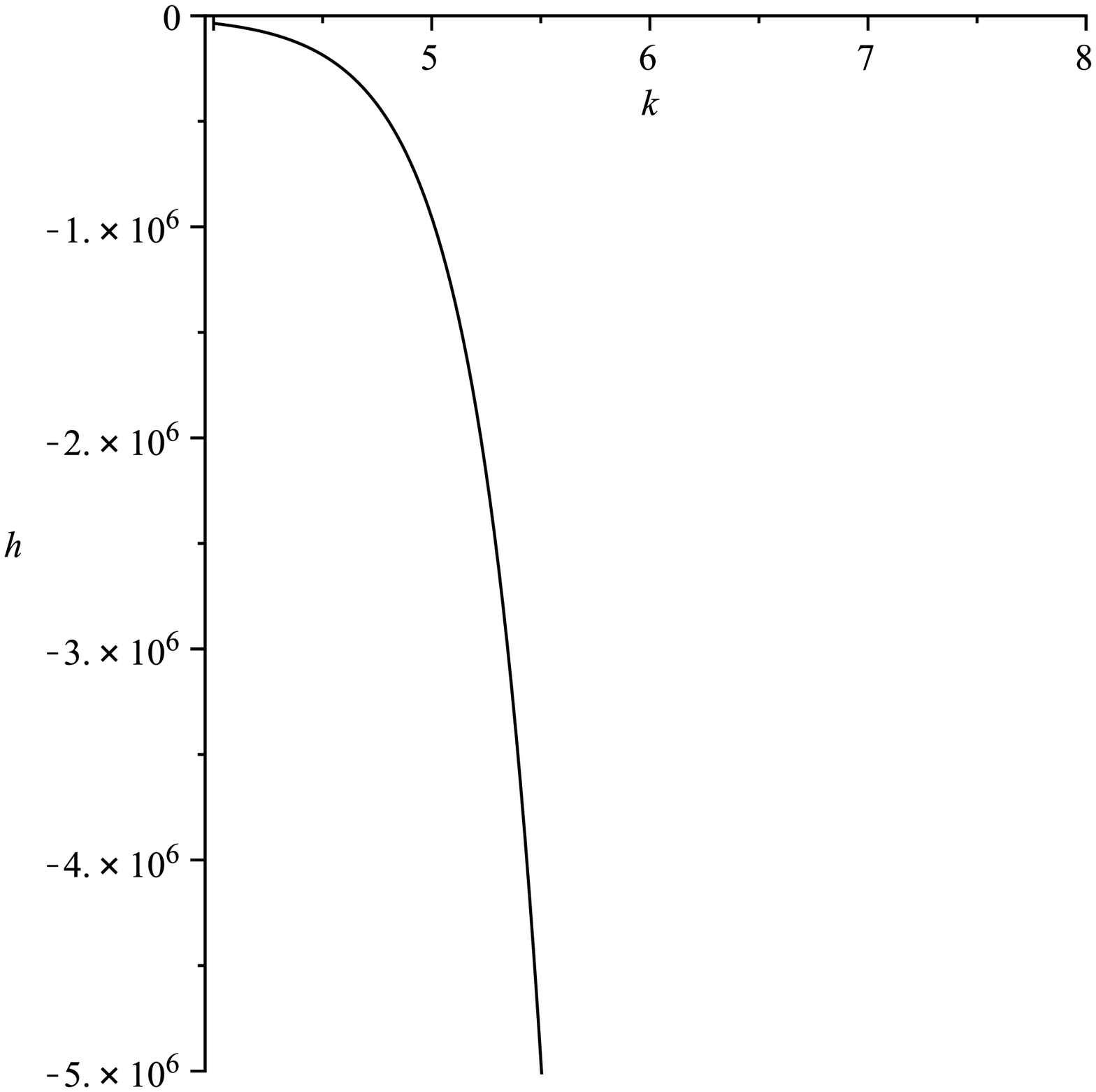} \ \ \ \includegraphics[width=6cm]{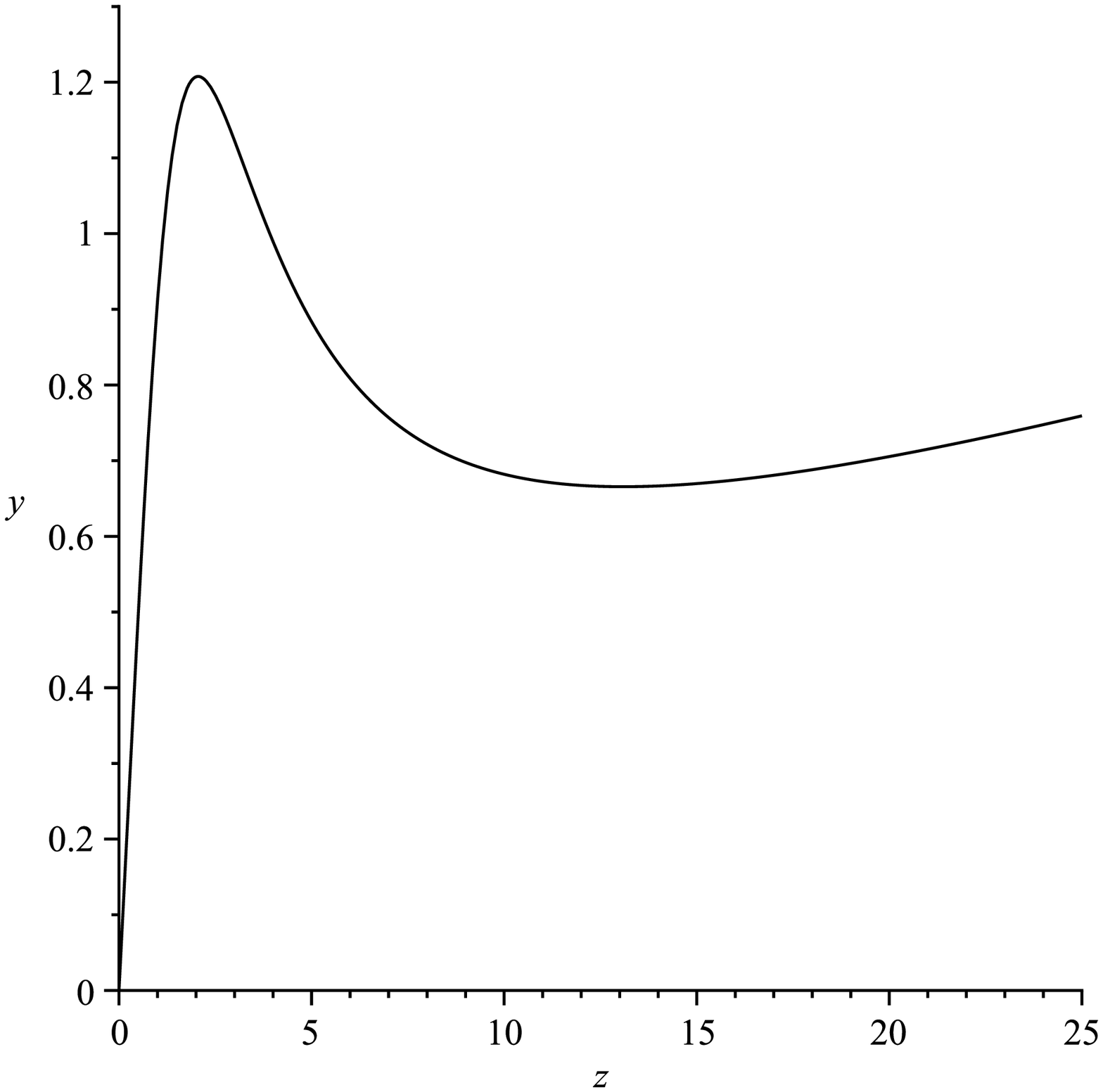}
\end{center}
\begin{center}{\footnotesize \noindent
 Рис.~2.
Графики функций $h(k+1)$ (слева) и $\varphi(z)$ при $k=6$ (справа)}.
\end{center}

\textbf{Замечание 3.} Воспользуясь Теоремой 3, можно найти $\lambda_{cr}^{(1)}\approx
0.6655887267, \ \lambda_{cr}^{(2)}\approx 1.207665883$ при $k=6$ и $\lambda_{cr}^{(1)}\approx
0.4661975987, \ \lambda_{cr}^{(2)}\approx 1.34764746$ при $k=7$.\

\subsection{Случай $G=\textit{ключ}$}\

Из (\ref{rus3.02}) можем получить ($z=z_1$)
$$z=\lambda\cdot \left({z \over\sqrt[k+1]{\lambda(2z+1)^k}}+1\right)^k=f(z),$$
производная правой части которого равна
$$f'(z)=k\lambda \left({z\over\sqrt[k+1]{\lambda(2z+1)^k}}+1\right)^{k-1}{2z+k+1\over \sqrt[k+1]{\lambda(2z+1)^{2k+1}}}.$$
Легко показать, что уравнение $z=f(z)$ имеет более одного решения
тогда и только тогда, когда уравнение $zf'(z)=f(z)$ имеет более
одного решения. Поэтому рассмотрим уравнение $zf'(z)=f(z)$, которое эквивалентно уравнению
$$\lambda={z^{k+1}[2(k-1)z+k^2+k-1]^{k+1}\over (2z+1)^{2k+1}}=\varphi_1(z).$$
Ясно, что $\varphi_1(0)=0$ и $\varphi_1(z)\rightarrow+\infty$ при $z\rightarrow+\infty$. Вычислим производную
$$\varphi_1^{'}(z)={z^k\cdot\left[2(k-1)z+k^2+k-1\right]^k\cdot\left[4(k-1)z^2-2(k^3-k^2-k+2)z+(k-1)(k^2+k-1)\right]\over (2z+1)^{2k+2}}.$$
Функция $\varphi_1(z)$ строго возрастает, если дискриминант $D(k)=4k^2(k^4-2k^3-5k^2+2k+5)$ квадратного трехчлена $4(k-1)z^2-2(k^3-k^2-k+2)z+(k-1)(k^2+k-1)$ отрицательный. Можно увидеть, что $D<0$ при $k=2$ и $k=3$. Значит, в этих случаях каждому значению $\lambda$ соответствует только одно значение $z$.
Следовательно, уравнение $z=f(z)$ имеет не более одного решения
при $k=2, \ k=3, \ \lambda>0.$

\begin{center}
\includegraphics[width=6cm]{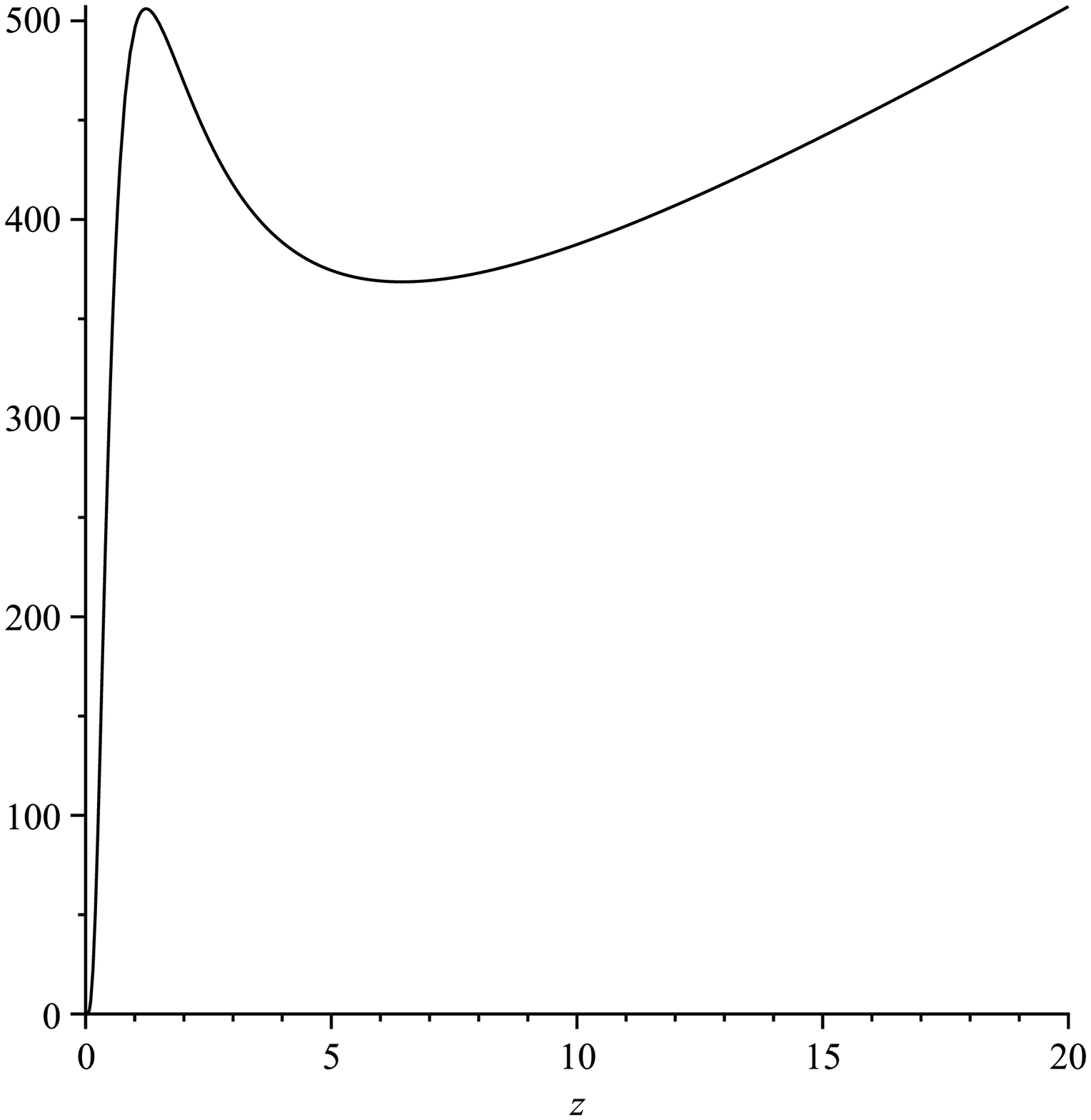}
\end{center}
\begin{center}{\footnotesize \noindent
 Рис.~3.
  График функции $\varphi_1(z)$ при $k=4$}.
\end{center}

Можно показать, что $D>0$ при $k\geq4$. Тогда анализ функции $\varphi_1(z)$ показывает, что она строго возрастает при $z<z_2, \ z>z_1$ и убывает при $z_2<z<z_1$, где $z_1, z_2$ ($z_1>z_2>0$) точки экстремума функции $\varphi_1(z)$ (Рис.3):
$$z_{1,2}={k^3-k^2-k+2 \pm k \sqrt{k^4-2k^3-5k^2+2k+5}\over 4(k-1)}.$$

Следовательно, уравнение $zf^{'}(z)=f(z)$ (соответственно, и уравнение $z=f(z)$) имеет одно решение при $\lambda>\lambda_{cr}^{(2)}, \ \lambda<\lambda_{cr}^{(1)}$ и имеет более одного решения при $\lambda_{cr}^{(1)}\leq\lambda\leq\lambda_{cr}^{(2)}$, где
\begin{equation}\label{rus3.7}
\lambda_{cr}^{(1)}=\varphi_1(z_1)=\left({k-1\over 2k}\right)^k\cdot{\left[(k^3-k^2-k+2+k\sqrt{D_1})\cdot(k^2+k+1+\sqrt{D_1})\right]^{k+1}\over 4\left(k^2-k+1+\sqrt{D_1}\right)^{2k+1}},
\end{equation}
\begin{equation}\label{rus3.8}
\lambda_{cr}^{(2)}=\varphi_1(z_2)=\left({k-1\over 2k}\right)^k\cdot{\left[(k^3-k^2-k+2-k\sqrt{D_1})\cdot(k^2+k+1-\sqrt{D_1})\right]^{k+1}\over 4\left(k^2-k+1-\sqrt{D_1}\right)^{2k+1}}.
\end{equation}
Здесь $D_1=k^4-2k^3-5k^2+2k+5$. Отсюда верна следующая

\textbf{Теорема 4.} \textit{Пусть $k\geq4$. Тогда в случае $G=\textit{ключ}$ для Hard-Core модели существуют $\lambda_{cr}^{(1)}, \ \lambda_{cr}^{(2)}$, определенные формулами (\ref{rus3.7}) и (\ref{rus3.8}), такие, что при $\lambda<\lambda_{cr}^{(1)}$ и $\lambda>\lambda_{cr}^{(2)}$ существует только одна трансляционно-инвариантная мера Гиббса, при $\lambda_{cr}^{(1)}\leq\lambda\leq\lambda_{cr}^{(2)}$ существуют более одной трансляционно-инвариантной меры Гиббса.}

\textbf{Замечание 4.} Заметим, что в работе \cite{KhR1} при $k=2$ и в работе \cite{XR2} при $k=3$ было доказано, что в случае $G=\textit{ключ}$ для любых $\lambda>0$  существует только одна Hard-Core трансляционно-инвариантная мера Гиббса.\

\end{document}